\documentclass[12pt]{article}

\usepackage[sc]{mathpazo}
\usepackage{chicago}
\usepackage[breaklinks=true]{hyperref}
\usepackage{breakcites}

\usepackage{amssymb,amsmath}                                           
\usepackage{bbm}                                                       
\usepackage{tensor}                                                    
\usepackage[usenames,dvipsnames]{color}                                
\usepackage{setspace}                                                  
\usepackage{hyperref}                                                  
\usepackage[left= 1.5in,right=1.5in,top=1in,bottom=1in]{geometry}          
\usepackage{comment}                                                   
\usepackage{authblk}                                                   
\usepackage{graphicx}                                                  
\usepackage{caption}                                                    
\usepackage{subcaption}

\hypersetup{                 
    colorlinks=black,         
    linkcolor=black,          
    citecolor=black,           
    urlcolor=black          
}

\let\oldmarginpar\marginpar
\renewcommand\marginpar[1]{\oldmarginpar{\color{red}\raggedright\scriptsize #1}}

\newcommand{\ket}[1]{\ensuremath{ \left| #1 \right\rangle }}

\title{Regarding the `Hole Argument' \\ and the `Problem of Time' }

\author[1]{\bf Sean Gryb\thanks{email: \href{mailto:s.gryb@hef.ru.nl}{s.gryb@hef.ru.nl}}}
\author[2]{\bf Karim P. Y. Th\'ebault\thanks{email: \href{mailto:karim.thebault@gmail.com}{karim.thebault@gmail.com}}}

\affil[1]{\small{Institute for Mathematics, Astrophysics and Particle Physics, Radboud University, Nijmegen, The Netherlands}}
\affil[2]{\small{Department of Philosophy, University of Bristol, United Kingdom}}

\date{\today}

\begin{document}
\sloppy
\maketitle

\begin{abstract}
The canonical formalism of general relativity affords a particularly interesting characterisation of the infamous hole argument. It also provides a natural formalism in which to relate the hole argument to the problem of time in classical and quantum gravity. In this paper we examine the connection between these two much discussed problems in the foundations of spacetime theory along two interrelated lines. First, from a formal perspective, we consider the extent to which the two problems can and cannot be precisely and distinctly characterised. Second, from a philosophical perspective, we consider the implications of various responses to the problems, with a particular focus upon the viability of a `deflationary' attitude to the relationalist/substantivalist debate regarding the ontology of spacetime. Conceptual and formal inadequacies within the representative language of canonical gravity will be shown to be at the heart of both the canonical hole argument and the problem of time. Interesting and fruitful work at the interface of physics and philosophy relates to the challenge of resolving such inadequacies.
 
\end{abstract}

\clearpage
\tableofcontents

\section{Introduction}

The `hole argument' is an interpretational dilemma in the foundations of general relativity. Although the argument originates with Einstein, the terms of the modern debate were set by \citeN{earman:1987}. In essence, Earman and Norton argue that a certain view on the ontology of spacetime (`substantivalism') leads to an underdetermination problem when pairs of solutions are connected by a certain sub-set of symmetries of the theory (`hole diffeomorphisms'). The dilemma is then either to give up substantivalism or accept that that there are distinct states of affairs which no possible observation could distinguish. The problem is supposed to derive purely from interpretational questions: it is assumed throughout that the formalism is rich enough to give an unambiguous representation of the relevant mathematical objects and maps -- it is just that different interpretations lead to different such choices.

The `problem of time' is a cluster of interpretational and formal issues in the foundations of general relativity relating to both the representation of time in the classical canonical formalism, and to the quantization of the theory. The problem was first noticed by Bergmann and Dirac in the late 1950s, and is still a topic of debate in contemporary physics \cite{Isham:1992,Kuchar:1991,anderson:2012}. Although there is not broad agreement about precisely what the problem is supposed to be, one plausible formulation of (at least one aspect of) the problem is in terms of a dilemma in which one must choose between an ontology without time, by treating the Hamiltonian constraints as purely gauge generating, and an underdetermination problem, by treating the Hamiltonian constraints as generating physical change \cite{pooley:2006,thebault:2012}. The problem derives, for the most part, from a formal deficiency in the representational language of the canonical theory: one cannot unambiguously specify refoliation symmetries as invariances of mathematical objects we define on phase space. We do not have a representation of refoliation symmetries in terms of a group action on phase space curves. 

What do these two problems have to do with each other? In this paper, we argue that, from a certain perspective, they are \textit{essentially the same problem} clothed in different language. The adoption of this perspective depends upon three key steps -- all of which are contestable -- but each of which is plausible. The first step requires accepting a certain view on the role of mathematics in guiding interpretational dispute within physical theory, and is inspired by a recent article by \citeN{weatherall:2015} (see \S2.1). The second step requires the adoption of the canonical formulation of general relativity as the primary forum for ontological debate regarding classical gravity (see 
\S2.2). This is motivated by the third step, whereby the ultimate goal of all debate regarding the ontology of classical spacetime is assumed to be the construction of a quantum theory of gravity (see \S3). Thus, methodologically speaking, much of what we say depends upon accepting a pragmatic view as regards the ontology of classical gravity and a canonical view as regards to the quantization of gravity. To those who will follow us this far -- we are grateful for the company.

\section{The Hole Argument}
 
\subsection{A Covariant Deflation}

In this section, we give a brief reconstruction of the aspects of the arguments of \citeN{weatherall:2015} that are relevant to our purpose. Although we are sympathetic to them, we will not argue explicitly in favour of Weatherall's negative conclusions regarding the cogency of the hole argument. Rather, we take his work to provide a clear means to differentiate two species of interpretational debate: one of significant pragmatic value, one of little. Let us begin with quoting some important passages of a more general methodological character: 
\begin{quote}
...the default sense of ``sameness" or ``equivalence" of mathematical models in physics \textit{should} be the sense of equivalence given by the mathematics used in formulating those models... mathematical models of a physical theory are only defined up to isomorphism, where the standard of isomorphism is given by the mathematical theory of whatever mathematical objects the theory takes as its models...isomorphic mathematical models in physics \textit{should} be taken to have the same representational capacities. By this I mean that if a particular mathematical model may be used to represent a given physical situation, then any isomorphic model may be used to represent that situation equally well. [pp.3-4 italics added] 
\end{quote}
Weatherall gives here a normative prescription regarding the kinds of debate that we \textit{should} have about the interpretation of physical theory. Mathematics provides us with a standard of equivalence in terms of an appropriate type of isomorphism. We should take this standard to dictate when mathematically models have identical representational capacities. Given that, arguments which depend upon interpreting isomorphic models as having different representational capacities are rendered ill-conceived. So far as it goes, this is a reasonable, if controversial, viewpoint. Our strategy here will be to provisionally accept such a view and see what happens to the debates regarding the hole argument and problem of time. We will return to the discussion of wider methodological morals in the final section. 

Following \citeN{weatherall:2015}, we can construct a version of the Earman and Norton hole argument as follows. Consider a relativistic spacetime as given by a Lorentzian manifold $(M,g_{\mu\nu})$.\footnote{For simplicity sake, throughout this paper we will take ourselves to be dealing with general relativity in vacuo. The Lorentzian manifolds in question will thus be solutions to the vacuum Einstein field equations. It is reasonable to assume that the arguments of this paper will apply, mutatis mutandis, to the matter case -- in which, of course, the `hole' would take on a more physical significance.} Consider a `hole' to be defined by some open proper set with compact closure $O \subset M$. A `hole diffeomorphism' is a map $\psi: M \rightarrow M$ with the properties that i) $\psi$ acts as the identity on $M-O$; and ii)  $\psi$ is not the identity on $O$. Given such a hole diffeomorphism we can define a metric tensor $\tilde{g}_{\mu\nu}$ in terms of the action of its push forward $\psi_\star$ on the original metric, i.e., $\tilde{g}_{\mu\nu}=\psi_\star(g_{\mu\nu})$. We have thus defined a transformation, $\tilde{\psi}: (M,g_{\mu\nu}) \rightarrow (M,\tilde{g}_{\mu\nu})$, producing a second relativistic spacetime which is \textit{isometric} to the first. Given that these are both Lorentzian manifolds and that isometry is the `standard of isomorphism' for Lorentzian manifolds, if we follow Weatherall's methodological prescription above, we should take $(M,g_{\mu\nu})$ and $(M,\tilde{g}_{\mu\nu})$ to have identical representational capacities. Yet, according to Weatherall, the crux of the `hole argument' depends upon there being a view as to the ontology of the theory within which $(M,g_{\mu\nu})$ and $(M,g'_{\mu\nu})$ have different representational capacities. In particular, it is assumed by Earman and Norton that  a `spacetime substantivalist' takes the two models to represent different assignments of the metric to points within $O$. This, according to Weatherall, is illegitimate since such a difference relies upon a comparison in terms of the identity map, $1_M:M\rightarrow M$, \textit{which is not the appropriate standard of isomorphism for the objects under consideration}. Moreover, the supposed dilemma rests upon conflating one sense of equivalence (in terms of $\tilde{\psi}$) with another sense of inequivalence (in terms of $1_M$ ):
\begin{quote}
...one cannot have it both ways. Insofar as one wants to claim that these Lorentzian manifolds are physically equivalent...one has to use $\tilde{\psi}$ to establish a standard of comparison between points. And relative to this standard, the two Lorentzian manifolds agree on the metric at every point--there is no ambiguity, and no indeterminism...Meanwhile, insofar as one wants to claim that these Lorentzian manifolds assign different values of the metric to each point, one must use a different standard of comparison. And relative to this standard---that given by $1_M$---the two Lorentzian manifolds are not equivalent. One way or the other, the Hole Argument seems to be blocked. [p.13]

\end{quote}

This is no doubt a rather controversial conclusion: Earman and Norton's hole argument has been the focus of debate in the philosophy of spacetime for almost thirty years. Could it really rely on such a simple misapplication of mathematics?  As noted above, the purpose of the present paper is not to enter into a sustained critical analysis of Weatherall's paper. Rather, we will briefly consider a plausible line of criticism and then move on the positive moral that we take both Weatherall and his critics to agree upon. 

In the definition of `same representational capacities' quoted above Weatherall could mean two subtly different things. In particular, when we say that \textit{either} of a pair of isomorphic models \textit{may} be used to represent a given situation \textit{equally well}, there is an ambiguity as to how restrictive we are being. We could mean that, taken in isolation, the two models can always represent any given situation equally well but allow that, taken together, they may represent different situations (for example, once the representational role of one model is fixed, the other could be taken to represent a different possibility). We could also mean that the two models must, in all contexts, always have to retain the ability to represent all physical scenarios equally well. If Weatherall means the former, then the hole argument is no longer `blocked'.\footnote{Thanks to Oliver Pooley (personal communication) for clarifying this.} If he means the latter, it is arguable that his notion of `same representational' capacities is too restrictive -- see \cite{roberts:2014}.

These plausible lines of criticism not withstanding, we think there are important lessons to be learned from considering such a `deflationary' response to the hole argument. In our view, debate about the ontology of physical theory is most fruitful and interesting when driven by representational ambiguity. If we accept that within a given domain there is a natural standard of mathematical equivalence, and that this standard is the appropriate guide to representational capacity, then the work left for interpretative philosophers of physics is only ever likely to be of marginal importance to the articulation and development of the theory. However, if there is not a natural standard of mathematical equivalence within the relevant domain, or there are reasons to believe that the standard (or standards) available are not good guides for representational capacity, then interpretative philosophy of physics gains an important role in the articulation and development of the theory. Whether or not Weatherall's arguments `block' the hole argument, we think it is reasonable to say that they highlight precisely the reasons why the hole argument is not seen as a topic of particular importance to contemporary physics.\footnote{We take this view to be broadly in the same sprit as Curiel's \citeyear{Curiel:2015} response to the hole argument: `diffeomorphic freedom in the presentation of relativistic spacetimes does not \textit{ipso facto} require philosophical elucidation, in so far as it in no way prevents us from focusing on and investigating what is of true physical relevance in systems that general relativity models' (p.11).}

\subsection{A Canonical Reinflation}

Following on from the discussion of Belot and Earman \citeyear{BelotEar:1999,BelotEar:2001} and the debate between Rickles \citeyear{rickles:2005,rickles:2006} and \citeN{pooley:2006}, we can consider a version of the hole argument reconstructed within the canonical or `ADM' \shortcite{ADM:1960,Arnowitt} formulation of general relativity.\footnote{The canonical formalism has its origin in the work of Paul Dirac and Peter Bergmann towards the construction of a quantum theory of gravity. Important early work can be found in \cite{Bergmann:1949} and \cite{Dirac:1950}, the crucial result was first given in \cite{Dirac:1958b}. According to \citeN{Salisbury:2012} the same Hamiltonian was obtained independently at about the same time by B. DeWitt and also by J. Anderson. Also see \cite{salisbury:2007, salisbury:2010} for an account of little-known early work due to L\'eon Rosenfeld.} Following the treatment of \citeN{Thiemann:2007}, the first step in constructing a (matter free) `3+1' space and time formalism is to make the assumption that the manifold $M$ has a topology $M\cong\mathbb{R}\times\sigma$ where $\sigma$ is a three-dimensional manifold that we will assume to be closed but have an otherwise arbitrary (differentiable) topology. Consider the diffeomorphism $X: \mathbb{R}\times\sigma \rightarrow M;  (t,x)\rightarrow X(t,x)$. Given this, we can define the \textit{foliation} of $M$ into hypersurfaces $\Sigma_{t}:=X_{t}(\sigma)$ where $t\in\mathbb{R}$ and $X_{t}:\sigma\rightarrow M$ is an \textit{embedding} defined by $X_{t}(x):=X(t,x)$ for the coordinates $x^{a}$ on $\sigma$. What we are interested in specifically is the foliation of a spacetime, $M$, into spacelike hypersurfaces, $\Sigma_{t}$, and, thus, we restrict ourselves to spacelike embeddings (this restriction is already implicit in our choice of topology for $M$) . Decomposing the Einstein--Hilbert action in terms of tensor fields defined upon the hypersurfaces $\Sigma$ and the coefficients used to parametrise the embedding (the lapse and shift below)  leads to a `3+1' Lagrangian formalism of general relativity. Recasting this into canonical terms gives us the ADM action:
 \begin{eqnarray}
S=\frac{1}{\kappa}\int_{\mathbb{R}}dt\int_{\sigma}d^{3}x\{\dot{q}_{ab}P^{ab}-[N^{a}H_{a}+|N|H]\}.
\end{eqnarray}
Here $\kappa=16\pi G$ (where G is Newton's constant and we assume units where $c=1$), $q_{ab}$ is a Riemannian metric tensor field on $\Sigma$, and $P^{ab}$ its canonical momenta defined by the usual Legendre transformation. $N$ and $N^{a}$ are multipliers called the lapse and shift. $H_{a}$ and $H$ are the momentum and Hamiltonian constraint functions of the form: 
    \begin{eqnarray}
H_{a}&:=&-2q_{ac}D_{b}P^{bc},\\
H&:=&\frac{\kappa}{\sqrt{det(q)}}[q_{ac}q_{bd}-\frac{1}{2}q_{ab}q_{cd}]P^{ab}P^{cd}-\sqrt{det(q)}\frac{R}{\kappa}.
\end{eqnarray}
It can be shown that, given a Lorentzian spacetime as represented by the geometry $(M,g_{\mu\nu})$, if the constraint equations are satisfied on every spacelike hypersurface, then $g_{\mu\nu}$ will also satisfy the vacuum Einstein field equations. Conversely, it can be shown that, given a $(M,g_{\mu\nu})$ that satisfies the vacuum Einstein field equations, the constraint equations, that are given by the weak vanishing of $H_{a}$ and $H$, will be satisfied on all spacelike hypersurfaces of $M$ (see  \cite{Isham:1992}  for details of both proofs). The solutions presented to us by the canonical and covariant formalisms are equivalent \textit{provided the covariant spacetime can be expressed as a sequence of space-like hypersurfaces}. This requirement is equivalent to insisting that the spacetimes in question are restricted to be globally hyperbolic \cite{Geroch:1970} and is directly connected to the topological restriction $M\cong\mathbb{R}\times\sigma$ that was made in setting up the canonical formalism. Thus, there is a subset of the covariant solutions that cannot be represented within the canonical formalism. 

There is a similar partial inequivalence between the formalisms at the level of symmetries, and again this difference relates to the topological restriction required to set up the 3+1 split. Whereas the covariant action is invariant under the full set of spacetime diffeomorphisms, Diff($M$), in the canonical formulation, it is only a subset of these transformations that is realised: those diffeomorphisms that preserve the spacelike nature of the  embedding.\footnote{The origin and nature of the difference between the symmetries of the canonical and covariant formalism is a complex issue. In addition to the restriction to diffeomorphisms that preserve the spacelike nature of the  embedding, the canonical formalism also neglects: i) field-dependent infinitesimal coordinate transformations; and ii) `large diffeomorphisms' that are not connected with the identity. See  \shortcite{Isham:1985a,Pons:1997,Pons:2010}.} We can examine this difference more carefully by considering the Lie  algebroid that the constraints generate:

\begin{eqnarray}
\{\vec{H}(\vec{N}),\vec{H}(\vec{N'})\}&=&-\kappa \vec{H}(\mathfrak{L}_{\vec{N}}\vec{N}') \label{eq:hda1},\\
\{\vec{H}(\vec{N}),H(N)\}&=&-\kappa H(\mathfrak{L}_{\vec{N}}N)\label{eq:hda2}, \\
\{H(N),H(N')\}&=&-\kappa \vec{H}(F(N,N',q))\label{eq:hda3},
\end{eqnarray}
where $H(N)$ and $\vec{H}(\vec{N})$ are smeared versions of the constraints (e.g. $\vec{H}(\vec{N}) :=\int_{\sigma}d^{3}xN^{a}H_{a}$) and $F(N,N',q)=q^{ab}(NN'_{,b}-N'N_{,b})$. The presence of structure functions on the right hand side of Equation \eqref{eq:hda3} is what prevents closure as an algebra, and means that the associated set of transformations on phase space are a groupoid rather than a group. This mathematical subtlety will be of great importance to our discussion. When the vacuum equations of motion and the constraints are satisfied, the groupoid of transformations generated by the constraints is equivalent to a subgroup of spacetime diffeomorphism group consisting of those diffeomorphisms that preserve the spacelike nature of the embedding and are connected to the identity.  

The `standard interpretation' of the representational roles of objects within the canonical formalism is as follows: a) Points on the constraint surface, defined by the sub-manifold of phase space where the constraints hold, represent Riemannian three geometries (together with the relevant momentum data); b) Integral curves of the Hamilton vector field of the Hamiltonian constraint on the constraint surface (dynamical curves for short) represent Lorentzian four geometries; c) Points connected by integral curves of vector fields associated with each family of constraints have identical representational capacities. There are problems with all three of these assignments of representational roles to objects within the canonical gravity formalism. In particular, with regard to a) and b), it is far from clear that points or curves should really be understood as playing such a simple role -- surely we need to consider the embedding data also? Furthermore, with regard to c), justification for the definition of equivalence classes is needed: this notion of `gauge orbits' derives from \textit{other} application of the theory of constrained Hamiltonian mechanics (see in particular \cite{Dirac:1964}) and we should require explicit physical reasons for its extension to the case of canonical gravity. Moreover, since the Hamiltonian of canonical general relativity is also a constraint, the integral curves that the standard interpretation implies should be identified as gauge orbits will also be identified as solutions! It will prove instructive to proceed as follows: we will assume a) and b) to be reasonable for the time being, and then investigate c) by considering the action of the constraints in the context of two canonical reconstructions of the hole argument. 

A first version of the `canonical hole argument', which seems to be largely what \citeN[\S2]{rickles:2005} has in mind,\footnote{Although Rickles' paper is mainly focused upon canonical gravity expressed in terms of connection variables, his reconstruction of the hole argument, like that considered here, relies solely upon the momentum constraints. We will consider Rickles' arguments further in the quantum context in \S3.2.} runs as follows. Consider a fixed foliation and define a single global Hamiltonian function that evolves the canonical data on three-geometries. Now consider a point on the constraint surface. This point corresponds to a particular specification of the metric tensor field, $q_{ab}$, and its associated canonical momenta $P^{ab}$. Given this, we can define a three-dimensional spatial geometry by the three dimensional Riemannian manifold $(\sigma,q_{ab})$. This object represents an instantaneous spatial slice of some physically possible spacetime. Because of the topological restriction, we know that such a slice is also a Cauchy surface, and thus can act as well-posed initial data for the spacetime in question (provided further smoothness conditions are satisfied). Explicitly, given this point, together with our global Hamiltonian, we can evolve the initial data backwards and forwards in time to get a path in phase space. We can then understand this path as representing an Einstein spacetime as a sequence of spatial slices. 

From this setup, we can construct an underdetermination problem analogous to the hole argument. Consider a transformation of canonical variables generated by the momentum constraint:
\begin{eqnarray}\label{eq:mom flowa}
\{\vec{H}(\vec{N}),q_{ab}\}=\kappa(\mathfrak{L}_{\vec{N}}q_{ab}) \label{eq:mom flowa},\\
\{\vec{H}(\vec{N}),P^{ab}\}=\kappa(\mathfrak{L}_{\vec{N}}P^{ab}) \label{eq:mom flowb}.
\end{eqnarray}  
The appearance of the Lie derivative on the right-hand side of each equation indicates that the momentum constraints can be understood as implementing the Lie group of (infinitesimal) diffeomorphisms of $\Sigma$ \cite{Isham:1985a}. This fits with the `standard interpretation' of the representational roles of objects within the canonical formalism. 
By acting on a point on the constraint surface with the momentum constraints (and appropriate smearing functions) we generate `flow' to a new point in phase space. This new point is associated with a second three dimensional Riemannian geometry $(\Sigma,\tilde{q}_{ab})$ that will be isometric to the first. A version of the canonical hole argument then follows from identifying canonical data within a subset of a phase space path constituting the `hole region' and comparing two curves $\gamma$ and $\tilde{\gamma}$ that differ solely in virtue of the action of the momentum constraints in that region. One could then argue that a substantivalist about \textit{space} would take the different canonical data within the hole region to correspond to sequences of spatial geometries, say $(\Sigma,q_{ab})_i$ and $(\Sigma,\tilde{q}_{ab})_i$, that represent different possibilities since they represent different assignments of the metric to spatial points on the slices. As pointed out by Pooley \citeyear{pooley:2006}, such a (rather na\"ive) spatial substantivalist position would involve commitment to a merely haecceitistic ontological difference: `These histories involve exactly the same sequence of geometrical relations being instantiated over time. The only way they differ is in terms of which points instantiate which properties'. By all accounts, however, the two curves $\gamma$ and $\tilde{\gamma}$, will represent the same spatial ontology outside the hole region. Thus, the spatial substantivalist is faced with the dilemma of either giving up their ontological view or submitting to underdetermination. 

So far things are looking very familiar. Essentially, all we have done is canonically reconstruct one very specific model of the covariant hole argument where the spacetime is constrained to be globally hyperbolic and the hole diffeomorphisms act only on spatial slices (this could be achieved given a particular choice of smearing function). It is therefore unsurprising that Weatherall's deflationary argument can be brought to bear upon this canonical version of the hole argument with equal force as upon its covariant cousin. Since each of the $(\Sigma,q_{ab})_i$ and $(\Sigma,\tilde{q}_{ab})_i$ are isometric as Riemannian manifolds, and isometry is the `standard of isomorphism' for Riemannian manifolds, we should take them to have identical representational capacities.  Moreover, the supposed ontological difference between the two spatial geometries relies upon a comparison in terms of the identity map $1_\Sigma:\Sigma\rightarrow \Sigma$, which is not the appropriate standard of isomorphism for the objects under consideration. Thus, we should not make interpretive arguments that rely upon the $(\Sigma,q_{ab})_i$ and $(\Sigma,\tilde{q}_{ab})_i$ representing distinct ontologies. To the extent that the covariant hole argument can be deflated in Weatherall's terms, so can this canonical version.  So far as the momentum constraints go, the `standard interpretation' of the representational roles of objects within the canonical formalism fits well with a Weatherall-style response to the hole argument.

However, following \citeN[\S2]{pooley:2006}, it is questionable whether the canonical hole argument should be given purely in terms of the action of the momentum constraints. After all, the restriction to a `fixed embedding' hole diffeomorphism is extremely strong: why should we not consider transformations that do not preserve the foliation? When thinking about the hole argument in the canonical formalism, one should surely try to reconstruct all the hole diffeomorphism that one can; i.e., one should include those which may deform the embedding so long as it remains space-like. To do this, one must also consider the transformations generated by the Hamiltonian constraints. This is where things become more difficult within the canonical formalism. In Equations \eqref{eq:mom flowa} and \eqref{eq:mom flowb} above, the connection between the momentum constraints and infinitesimal diffeomorphisms is made explicit by the occurrence of the Lie derivatives of the canonical variables in the direction defined by the shift multiplier, $\mathfrak{L}_{\vec{N}}q_{ab}$ and $\mathfrak{L}_{\vec{N}}P^{ab}$. The form of these expressions indicate that the phase-space action of $\vec{H}(\vec{N})$ can be associated with diffeomorphisms of the original spacetime manifold, $M$, tangential to the embedded hypersurfaces $\Sigma_{t}$. Clearly, this does not exhaust the set of possible diffeomorphisms that can be represented within the canonical formalism since we may also consider diffeomorphisms of $M$ that are orthogonal to  $\Sigma_{t}$ -- these would be the `time bit' of the spacetime diffeomorphism group, as opposed to the `space bit' . In for us to represent the full set of the canonical symmetries of the theory, we might therefore hope that the Hamiltonian constraints can be associated with an action of the form: `$\{H(N),q_{ab}\}=\kappa(\mathfrak{L}_{Nn}q_{ab})$', where $n^{\mu}$ is the unit normal vector to $\Sigma_{t}$, and $n=n^{\mu}$ . However, such an equation \textit{is not} found in explicit calculation -- see \cite{Thiemann:2007} Eq. (1.3.4) and (1.3.12). Rather, what is found is that, in the case of the metric variable $q_{ab}$, the expected $\mathfrak{L}_{Nn}q_{ab}$ piece emerges only `on shell' --  i.e., only when the equations of motion hold -- and relative to an embedding. We therefore have that, whereas the diffeomorphisms associated with the momentum constraints can be understood as purely kinematical symmetries of the three geometries $\Sigma$ (irrespective of whether the equations of motion hold), those associated with the Hamiltonian constraint are properly considered symmetries of, not only entire spacetimes, but of spacetimes which are solutions. (We will return to the problem of foliation symmetry and the Hamiltonian constraints in \S3.1).

This creates an immediate problem for the  `standard interpretation' of the representational roles of objects within the canonical formalism. In particular, unlike in the case of momentum constraints, for the case of the Hamiltonian constraints, the explicit form of the constraint's phase space action does not justify an interpretation in which points connected by integral curves of vector fields associated with constraints have identical representational capacities -- i.e., c) above is now unsupported. 

Essentially, the problem is that, in the  `standard interpretation', the objects that represent spacetimes are dynamical curves on the constraint surface, and for curves related by the action of the Hamiltonian constraints we do not have a standard of isomorphism that is given by the mathematical theory of the objects at hand: i.e. phase space curves. This is because the Hamiltonian constraints do provide us with anything like a representation of refoliation symmetries as a group action on phase space curves. This issue is notwithstanding our ability to reconstruct the embedding and define a refoliation based upon phase space data. That is, given phase space data, and provided we are `on shell', we \textit{can} reconstruct the lapse and shift multipliers by solving the `thin sandwich problem' \cite{giulini:1999}. From there, we are able to define the embedding and then, as indicated above, unambiguously specifiy a refoliation transformation. However, that the phase space data are a \textit{sufficient starting point} to define a mapping between phase curves representing spacetimes related by a refoliation, does not automatically mean that we have an available a \textit{standard of isomorphism} between the phase space curves in question. In particular, the structure that is preserved  in a refoliation is encoded in the metric of the dynamical spacetimes in question, and is not contained in the relevant canonical data on their own. Based upon the data, we can reconstruct the embedding, and based upon that we can specify the class of mappings that deform the embedding whilst persevering the relevant spacetime metric (and topological) structure. The case of refoliations in canonical gravity is thus crucially different from spatial diffeomorphisms: refoliations are not invariances of mathematical objects we define on phase space. This is true even if we can reconstruct the mathematical objects that they are invariances of based upon the phase space objects.    

We can still, however, determine, given two dynamical curves, when they are \textit{relatable} by a refoliation, based purely upon the geometric structure of phase space. Consider the surface, $\Pi$, within the phase space $(q_{ab},P^{ab})\in \Gamma$, defined by satisfaction of the Hamiltonian constraint equations $H=0$. The integral curves of the vector fields associated with the Hamiltonian constraint define sub-manifolds within $\Pi$. Formally speaking, they \textit{foliate} the presymplectic manifold $\Pi$ into symplectic sub-manifolds, $L$, that are the \textit{leaves} of the foliation. The geometry of the situation is then such that any two curves that lie within the same leaf will be relatable by some possible refoliation transformation, given the appropriate supplemental data regarding the embedding of the two sequences of hyper-surfaces into a spacetime. Thus, although at the level of phase space we do not have an isomorphism between curves, we do still have a means, at the level of phase space, by which we can say that they are constrained to represent objects that are observationally indistinguishable.    

We can now consider a  `pure Hamiltonian constraint' hole argument. Consider a pair of canonical solutions in terms of two dynamical curves in phase space. Each of these curves represents a sequence of three geometries that we can then further take to represent a  spacetime. If these two curves are integral curves of the vector fields associated with the Hamiltonian constraints \textit{and} overlap for some nontrivial section \textit{and} both lie entirely inside a particular leaf $L$, then we can reconstruct a `pure Hamiltonian constraint' hole argument where we take the momentum constraints to have a trivial action our two curves, but still can consider an underdetermination scenario. One might think, in these circumstances, that Weatherall's deflationary strategy might be able to block this `pure Hamiltonian constraint' hole argument too. However, things are not quite so easy this time. The mathematical objects that we are using to represent spacetimes are curves in phase space and not four dimensional Lorenizan manifolds. In order to actually reconstruct a four dimensional Lorenizan manifold from a phase space curve, one is required to specify embedding data. And so long as we are talking about representation of spacetimes via the pure phase space formalism, there is no readily available mathematical standard of isomorphism. Whereas isometry between Riemannian three manifolds leads naturally to treating points connected by the momentum constraints as having the same representational capacity, there is no analogous \textit{mathematical} prescription for curves. We have no representation of refoliation symmetries as a group action on phase space curves.

We can consider a position of `straightforward'  substantivalism where spacetime points have some basic status such that different embeddings of spatial slices into spacetime would then correspond to different physical possibilities. A hole argument could then be reconstructed along Earman and Norton's lines, since we could have underdetermination given an appropriate class of embedding non-preserving hole diffeomophisms.  The usual options for a `sophisticated' version of substantivalism would then be back on the table and one could, if one is so inclined, pursue a response to the hole argument in terms of anti-Haecceitism,\footnote{This strategy is adopted in, for example, \cite{maudlin:1988,butterfield:1989,brighouse:1994}. See \cite{pooley:2013} for more details.} or perhaps some other option. What is more, in the context of this pure Hamiltonian constraint canonical hole argument, the spectre of underdetermination looms over both relationalist and substantivalist interpretations alike. In particular, as pointed out by \citeN{Pooley:2001}, in the context of the canonical formalism, Machian relationalism \textit{about time} is also confronted by indeterminism, unless some other steps are taken -- see also \cite{thebault:2012}. Our point here is not to reopen these debates, nor to blunt Weatherall's attack on the hole argument per se. Rather than resurrecting the hole argument with new splendour, we should seek to find a formalism fit to deflate it again! From this perspective, the canonical formalism is deficient precisely because it does not provide us with a natural standard of isomorphism for paths in phase space. The role of interpretational work is then to aid us in finding an adequate formalism, not to drive different articulations of the formalism once it is found. 

The two key conclusions from our analysis thus far are as follows: i) there is a version of the hole argument within the canonical formalism that is still well defined despite the \textit{failure} to find an appropriate  standard of isomorphism for the mathematical objects in question (paths in phase space); and ii) this problem essentially derives from the ambiguity regarding the canonical representation of diffeomophisms that do not preserve the embedding of spatial hypersurfaces into spacetimes. We have no representation of refoliation symmetries as a group action on phase space curves. On our reading, this is simply a restatement of one aspect of the problem of time, to which we now turn.

\section{The Problem of Time}

\subsection{The Problem of Refoliation}

Controversy surrounding the problem of time relates as much to its origin as to the purported solutions. On one influential view\footnote{Espoused in particular by \citeN{Rovelli:2004} but also echoed in much of the philosophical literature \cite{BelotEar:1999,BelotEar:2001,Earman:2002,Belot:2007}. Dissent from this `received view', for various reasons, can be found in \cite{BarbourI,Kuchar:1991,Maudlin:2002,Pons:2005,Thebault:2011b,thebault:2012,gryb:2011,gryb:2014, Gryb:2015,Gryb:2015a,Pons:2010,pitts:2014a,pitts:2014}.}, the problem is inherent in the application of Dirac's theory of constrained Hamiltonian mechanics, in which the first class constraints generate `unphysical' gauge transformations, to a theory in which the Hamiltonian is defined in terms of first class constraints. Most pertinently, since it is the Hamiltonian constraints that generate the evolution between dynamically related hypersurfaces, if these constraints are understood as `gauge generating', then we can jump straight to the somewhat paradoxical conclusion that `time is gauge'. However, it is not difficult to see that such a reading is not well supported. For one thing, there are good reasons to doubt that the theorem upon which the statement that `first class constraints generate unphysical gauge transformations' is based applies to theories with Hamiltonian constraints \cite{Barbour:2008}. For another, as we have already seen, the connection between spacetime diffeomorphisms and the Hamiltonian constraints is not a direct or simple one. In fact, if anything, the problem of time stems from the problem that we \textit{cannot} write an evolution equation using the phase space action of Hamiltonian constraints that would imply either dynamics or `unphysical' change. The problem derives from the idea that we should be able to represent a refoliation as an invariance of objects defined purely within phase space, and thus it is worth considering what we mean by a refoliation symmetry in a little more detail. 

The infinite family of Hamiltonian constraints in canonical general relativity is connected to \textit{local time-diffeomorphism invariance}. The set of local time-diffeomorphisms includes two very different types of transformations that are important to distinguish.  Canonical general relativity involves specification of geometrical information relating to both canonical data on sequences of spatial hypersurfaces and the embeddings of these hypersurfaces into spacetimes. Local time-diffeomorphisms are a set of symmetry transformations that includes both reparametrization transformations that preserve embeddings and refoliation transformations that do not preserve embeddings. The embedding preserving transformations reparametrize phase space curves without changing their image. This means that the sense in which reparametrizations can lead to a potential `hole argument' is a very weak one: one can have underdetermination in the sense of multiple substantivalist ontologies of time being compatible with an initial specification, but one does not have indeterminism in the sense of these ontologies corresponding to different phase space curves -- see \cite[\S3.1]{Gryb:2015} for more discussion of this point. 

Refoliations, on the other hand, change the image of phase space curves and so can lead to underdetermination problems with ontologies corresponding to different phase space curves. As we have already seen, such underdetermination problems are not susceptible to a Weatherall-style deflation precisely because there is not an appropriate standard of isomorphism between the relevant mathematical objects defined on phase space. At the level of phase space, we do not even have a unique state-by-state representation of refoliations. This is because, in addition to a history, one needs to specify the embedding of the phase space data into spacetime in order to define a refoliation.  Without this embedding information, it is impossible to construct the explicit refoliation map between two histories on phase space. Another way of viewing this issue is to consider refoliations as normal deformations of three-dimensional hypersurfaces embedded within four geometries \cite{Teitelboim:1973}. In that context, it is clear that they will require a spacetime metric in order to be defined. This metric can be defined either explicitly as spacetime geometric data or implicitly via reconstruction of the embedding from canonical data (via solution of the thin sandwich problem). 

There are at least five natural responses to the problem of refoliation: I) Move to a reduced formalism; II) Use internal clocks; III) Fall back on the covariant formalism; IV) Add new physical ingredients; and V) Add new mathematical ingredients. All of these responses have potentially significant implications for the representational roles of objects within the canonical formalism, and we will consider them briefly here with particular reference to the discussion of the `standard view' and the `reinflation' of the hole argument considered above. 

The first, reductive response is the standard response in the literature -- see in particular \cite{Belot:2007}. It is often (perhaps understandably) conflated with the second. The idea is that the integral curves of the vector fields on the constraint surface associated with the Hamiltonian constraints should be `quotiented out' to form a reduced phase space. Points within this reduced phase space are then taken to represent diffeomorphism invariant spacetimes since there is an  bijection between points in the reduced phase space and points in a space of diffeomorphism invariant spacetimes, defined via the covariant formalism. However, as discussed in \cite{thebault:2012}, the existence of such a mapping between points in two representative spaces is far from a sufficient condition for them to play equivalent roles (although it could in some cases be taken to be necessary) since we can trivially find such relationships between manifestly inequivalent structures. It is much more plausible for the representational role of a space within a theory to be fixed primarily by its relationship to the representative structures from which it is derived rather than to a space utilised in the context of a different formalism. For the case of general relativity, therefore, it is more appropriate to consider the relationship between the reduced phase space and the unreduced phase space as fixing the former's representational role. And so, we are back at the problem of interpreting points and curves in the unreduced phase space -- without first fixing such an interpretation, the representative role of the reduced phase space should be taken to be undefined. 

The second response involves using a series of internal clocks to (arbitrarily) parametrize the foliation. Correlations between the clock values are then used to construct `complete observables' \cite{Rovelli:2002,Dittrich:2007,Dittrich:2006}. There are problems with monotonicity and chaos \shortcite{Dittrich:2015aa} and it is not clear these can be overcome in general. Furthermore, the representational role of the clock values is contested: Dittrich and Thiemann \citeyear{Thiemann:2007,dittrich:2009} claim that only the correlations represent physical quantities,  Rovelli \citeyear{Rovelli:2002,Rovelli:2014} claims that the values themselves (partial observables) represent quantities that can be `measured but not predicted'. The Dittrich-Thiemann view is motivated by something like the standard interpretation, and is also closely allied with the reduced view: By treating only the correlations as physically meaningful one once more endorses an equivalence in representational capacity between a point on the initial data surface and an entire history. However, the different foliations are encoded in the clock values, not the correlations. We, thus, consider that it is only on the Rovelli view that the problem of refoliation could, at least in principle, be resolved via the use of internal clocks. See \cite{Gryb:2015a} for further discussion of the relative merits of the Dittrich-Thiemann and Rovelli views. 

In the third response we simply rely upon correspondence to the covariant formalism as our guide to determine which phase space curves should be taken to represent spacetimes related by refoliations.  In following such an approach, we would use the covariant formalism as the primary guide to representational capacities, and thus strip the canonical formalism of any independent representational capacity. In this context, the deflationary argument of Weatherall should be re-applicable: i.e., we should be able to use the isometry between Lorentzian spacetimes in the covariant formalism to fix the representational capacities of the phase space curves in the canonical formalism related by refoliation. In spirit, such an approach appears to be close to what \shortciteN{Pons:2010}  and \citeN{pitts:2014} have in mind. It is also closely connected to approaches to the quantization of gravity based upon the path integral, for example causal set theory \shortcite{bombelli:1987,dowker:2005,henson:2006}, causal dynamical triangulation \shortcite{loll:2001,ambjorn:2001}, spin foams \cite{baez:1998,perez:2013}, or functional RG approaches \cite{lauscher:2001}.

In the fourth and fifth responses one admits that the canonical formalism is deficient precisely because it does not provide us with a natural standard of isomorphism for paths in phase space. However, rather than falling back on the covariant formalism one seeks to enrich the canonical theory by including either new physical ingredients (IV) or new mathematical ingredients (V). 

It is in the context of the fourth response that one can view the `Shape Dynamics' formalism \shortcite{Gomes:2011} of canonical gravity that was originally motivated by the project of implementing the `Machian program' for understanding space, shape and time in general relativity \shortcite{barbour:scale_inv_particles,barbour_el_al:scale_inv_gravity,barbour_el_al:physical_dof}. The relationship between Shape Dynamics and the problem of time is investigated in a series of papers by Gryb and Th\'{e}bault  \citeyear{gryb:2011,gryb:2014,Gryb:2015,Gryb:2015a}  and is also considered (along a different line) in \shortciteN{barbour:2014}. The crucial idea in both treatments is that, in Shape Dynamics, refoliations are `re-encoded' as conformal transformations, and the refoliation aspect of the problem of time is eliminated. More precisely, one moves to a new formalism `dual' to the sector of canonical gravity where the spacetimes are foliable by hypersurfaces of constant mean curvature. The constraints of this new space are a single Hamiltonian constraint (responsible for reparameterization) together with the usual momentum constraints, and a new set of constraints that generate (volume persevering) three-dimensional conformal transformations. The momentum and conformal constraints are amenable to the `standard interpretation' in that it is appropriate to view points connected by integral curves of vector fields associated with these constraints as having identical representational capacities. Such an interpretation is still inappropriate for the single Hamiltonian constraint since it would eliminate time evolution. However, as mentioned above, reparametrizations are  `embedding preserving' and simply relabel phase space curves without changing their image. If we accept that the standard of isomorphism for phase space curves is given by the mathematical theory of curves, then transformations that do not change the image of the curve clearly should be classified as isomorphic. In this context, curves related by reparametrizations can be taken to have the same representational capacities and even the shadow of a remaining hole argument can be deflated \`a la Weatherall. Although interpretational work was crucial in finding the Shape Dynamics solution to the problem of refoliation, once such a representationally adequate formalism has been constructed, there should not remain substantive debates as to the representational capacities of the relevant mathematical objects.    

The fifth response to the problem of refoliations is to stay within canonical general relativity and add in extra mathematical, rather than physical, ingredients. Such an approach would necessarily involve moving away from thinking about Lie group actions as relating classes of objects with identical representational capacities. Rather, the symmetries would have to be represented in terms of Lie groupoids and we should look to represent the groupoid of diffeomorphisms between space-like embeddings of hyper-surfaces within Lorentzian manifolds. Although some work is this direction was attempted by Isham and Kuchar \citeyear{Isham:1985a,Isham:1985b}, it is only relatively recently that this (rather fearsome) mathematical challenge has been addressed in earnest. In particular, \shortciteN{blohmann:2010} argue that the Poisson bracket relations among the initial value constraints for the Einstein evolution equations correspond to those among the constant sections of a Lie algebroid over the infinite jets of paths in the space of Riemannian metrics on a manifold. The goal of this project is to reconstruct the canonical formalism such that the constraint equations may be seen as the vanishing of something like a momentum map for a groupoid of symmetries. In such circumstances, the isomorphisms relevant for refoliation would be explicitly constructed, and one would expect Weatherall's deflationary strategy to be available. However, it remains to be seen whether and how this project will be completed. For the time being, we can simply note that the representational ambiguity that enables us to `reinflate' the hole argument is closely related to that which causes the problem of time, and is also an area of current research in the mathematical foundations of canonical gravity. This `problem of refoliation' is also of great importance to the quantization of gravity. In the next section, we will consider the relation between the problem of time (and the hole argument) in the context of the `problem of quantization', before concluding with some methodological morals.

\subsection{The Problem of Quantization}

If the problem of time were simply the problem of refoliation then, although it would perhaps still be a problem of interest to contemporary physicists, the problem of time would surely not be a problem of real consequence. The principal goal of work towards the canonical reformulation of general relativity was always to enable a quantization of the theory. For within the canonical formulation of a classical theory, we are accustomed to see the seeds of a quantum theory in terms of the Lie algebra of observables induced by the symplectic structure of phase space. Since canonical general relativity is a \textit{constrained} phase space theory, one seeks to apply a version of canonical quantization designed for such constrained Hamiltonian theories. This is the Dirac constraint quantization procedure \cite{Dirac:1964,Henneaux:1992a}. Setting aside a legion of technical issues, if one (rather heuristically) presses ahead and applies Dirac constraint quantization to canonical general relativity, one runs directly into a conceptual problem. Following `Dirac quantization', one follows a procedure that involves: i) first promoting the first-class constraints of the classical constrained Hamiltonian theory to operators acting upon a \textit{kinematical Hilbert space}, $\mathcal{H}_{kin}$ ; and then ii) imposing the constraint operators as restrictions on physically possible states, and in doing so constructing a \textit{physical Hilbert space}, $\mathcal{H}_{phys}$. When this procedure is applied to Hamiltonian constraints, an immediate conceptual worry arises since, by definition, the only physical states now permitted will be energy eigenstates. Applying Dirac constraint quantization to canonical general relativity leads to a `frozen formalism' with all physical information supposedly encoded in a Wheeler-DeWitt equation of the form `$\hat{H}\ket \Psi =0$'.  

So far as it goes, this story is rather incomplete. Neither the kinematical Hilbert space, nor the physical Hilbert space, nor the Wheeler-deWitt equation expressed in these terms are well-defined mathematical objects, and the technical challenge of rigorously enacting a constraint quantization of canonical general relativity is a significant one \cite{Thiemann:2007}. Although our purpose here is not to review the relevant technicalities in detail, there is one point of particular relevance to our argument. We saw in Equation \eqref{eq:hda3} that the Poisson bracket of the Hamiltonian constraints with themselves close with structure functions. This is what prevents closure of the constraints as an algebra, and means that the associated set of transformations on phase space are a groupoid rather than a group. This is also what blocks application of the more sophisticated modern cousin of Dirac constraint quantization: the procedure of `Refined Algebraic Quantization' (RAQ) \cite{Giulini:1999a} that is applicable to the momentum but not Hamiltonian constraints. In the RAQ approach one defines the physical states by `group averaging' $\Psi$ over the manifold associated with the Lie algebra of the constraints. This involves constructing a quantum mechanical analogue to the classical gauge orbits within the kinematical Hilbert space $\mathcal{H}_{kin}$ \cite{Corichi:2008}. Any two points along these `quantum gauge orbits'  yield the same physical state and so, in group averaging, we are explicitly removing kinematical redundancy at the level of states. It is at this point that we can note, against \citeN{rickles:2005} and in agreement with \citeN{pooley:2006}, that there cannot be a quantum analogue to the hole argument driven by the momentum constraints. The equivalence classes are not made up of physical states, and one could not, even in principle, consider solutions which contain different representatives of the same physical state. In constructing the physical Hilbert space we removed precisely the kinematical redundancy that makes the classical hole argument via the momentum constraints possible. This should perhaps be no surprise: a reliable quantization procedure should be expected to identify objects that are equivalent up to the relevant notion of isomorphism in the classical theory.\footnote{At least to the extent that we expect quantization to eliminate `surplus structure' encoded in the local symmetries of a gauge theory. See \cite{gryb:2014}. }      

This brings us to our central point. When we do not have an appropriate standard of isomorphism between a group of mathematical objects in our classical formalism, we have no good guide to the construction of an appropriate quantum theory. Thus, the problem of refoliation, wherein the formalism of canonical gravity does not provide a natural standard of isomorphism between phase space curves related by refoliation, becomes a problem of quantization, wherein we do not have a reliable quantization prescription for removing the kinematical redundancy relating to refoliations. Although there are quantization techniques that are formally applicable to Hamiltonian constraints -- for example the master constraint program \cite{Thiemann:2006} -- these all inevitably lead to a timeless formalism, with the universe trapped in an energy eigenstate. Moreover, within such approaches, no means is available to represent refoliations within a kinematical Hilbert space, and thus it seems questionable whether the right redundancy is being disposed of. In a sense, the quantum aspect of the problem of time exists precisely for the same reason that the canonical hole argument is resistant to deflation: refoliations do not admit a representation as a group of transformations on phase space. Thus, in trying to find a solution to the problem of refoliations, we will also be working towards solving the problem of quantizing gravity.       

\section{Methodological Morals}

In this paper, we hope to have illustrated two methodological morals that we believe philosophers of physics would do well to heed. First, interpretational debate regarding the foundations of a physical theory would do well to track ambiguity regarding the representational capacity of mathematical objects within that theory. Whilst we could always simply resist the normative force of Weatherall's  demand that `isomorphic mathematical models in physics should be taken to have the same representational capacities' \cite[p.4]{weatherall:2015}, we think the hole argument and the problem of time demonstrate that fruitful and exciting interpretive work often lies in the mathematically ambiguous terrain that is beyond its scope. When we have no natural standard of isomorphism between mathematical objects in a physical theory, debate regarding the representational capacity of these objects is of immediate importance for the articulation of the theory. 

Along similar lines, our second moral is that interpretational debate should track problems with some connection to theory articulation and development.\footnote{Again, we take ourselves to be partially working in the same sprit as Curiel \citeyear{Curiel:2015}. In particular, we entirely support the claim that: `metaphysical argumentation abstracted from the pragmatics of the scientific enterprise as we know it...is vain' (p.6).} The representational capacity of objects within a theory is an issue of real importance to the extent to which it potentially bears upon the articulation and development of a theory. In these terms, the debate about the ontology of classical spacetime is most interesting and important to the extent to which it has relevance to the pursuit of a quantum theory of gravity. By considering the hole argument in the context of the problem of time we see that the debate \textit{can} have such a bearing. 

\section*{Acknowledgements} 

We are hugely indebted to Erik Curiel, Samuel Fletcher, Oliver Pooley, Bryan Roberts and James Weatherall for discussion and written comments that were invaluable in the evolution of this paper. We are also very appreciative of feedback from members of audiences in Berlin, Bristol, London and Oxford. 

KT acknowledges the support of the Munich Center for Mathematical Philosophy, the Alexander von Humboldt foundation, and the University of Bristol. SG's work was supported by the Netherlands Organisation for Scientific Research (NWO) (Project No. 620.01.784).

\bibliographystyle{chicago}
\bibliography{Masterbib,Mach}

\end{document}